\documentclass[conference]{IEEEtran}

\usepackage{cite}
\usepackage{amsmath,amssymb,amsfonts}
\usepackage[pdftex]{graphicx}
\graphicspath{{images/}}
\usepackage{textcomp}
\usepackage{xcolor}
\usepackage{algorithm}
\usepackage{algpseudocode}
\usepackage{booktabs}
\usepackage{comment}

\def\BibTeX{{\rm B\kern-.05em{\sc i\kern-.025em b}\kern-.08em
    T\kern-.1667em\lower.7ex\hbox{E}\kern-.125emX}}
\begin{document}

\title{Optimizing Product Deduplication in E-Commerce with Multimodal Embeddings}

\author{
    \IEEEauthorblockN{Aysenur Kulunk\textsuperscript{\textsection}}
    \IEEEauthorblockA{\textit{Data Analytics Department} \\
    \textit{Hepsiburada}\\
    Istanbul, Turkey \\
    aysenur.kulunk@hepsiburada.com}
    \and
    \IEEEauthorblockN{Berk Taskin\textsuperscript{\textsection}}
    \IEEEauthorblockA{\textit{Data Analytics Department} \\
    \textit{Hepsiburada}\\
    Istanbul, Turkey \\
    berk.taskin@hepsiburada.com}
    \and
    \IEEEauthorblockN{M. Furkan Eseoglu\textsuperscript{\textsection}}
    \IEEEauthorblockA{\textit{Data Analytics Department} \\
    \textit{Hepsiburada}\\
    Istanbul, Turkey \\
    furkan.eseoglu@hepsiburada.com}
    \and
    \IEEEauthorblockN{H. Bahadir Sahin}
    \IEEEauthorblockA{\textit{Data Analytics Department} \\
    \textit{Hepsiburada}\\
    Istanbul, Turkey \\
    hikmet.sahin@hepsiburada.com}
}

\maketitle

\begingroup\renewcommand\thefootnote{\textsection}
\footnotetext{Equal contribution}
\endgroup

\begin{abstract}
In large scale e-commerce marketplaces, duplicate product listings frequently cause consumer confusion and operational inefficiencies, degrading trust on the platform and increasing costs. Traditional keyword-based search methodologies falter in accurately identifying duplicates due to their reliance on exact textual matches, neglecting semantic similarities inherent in product titles. To address these challenges, we introduce a scalable, multimodal product deduplication designed specifically for the e-commerce domain. Our approach employs a domain-specific text model grounded in BERT architecture in conjunction with MaskedAutoEncoders for image representations. Both of these architectures are augmented with dimensionality reduction techniques to produce compact 128-dimensional embeddings without significant information loss. Complementing this, we also developed a novel decider model that leverages both text and image vectors. By integrating these feature extraction mechanisms with Milvus, an optimized vector database, our system can facilitate efficient and high-precision similarity searches across extensive product catalogs exceeding 200 million items with just 100GB of system RAM consumption. Empirical evaluations demonstrate that our matching system achieves a macro-average F1 score of 0.90, outperforming third-party solutions which attain an F1 score of 0.83. Our findings show the potential of combining domain-specific adaptations with state-of-the-art machine learning techniques to mitigate duplicate listings in large-scale e-commerce environments.
\end{abstract}


\section{Introduction}
In today's vast e-commerce marketplaces, particularly within the Turkish e-commerce landscape, customers frequently encounter duplicate product listings that create confusion and frustration during shopping. These duplicates make it difficult to compare options, slow down decision-making, and diminish trust in the platform. From an operational standpoint, maintaining a catalog with duplicate products creates significant overhead—increased storage and computational costs for search systems. Moreover, duplicate listings can disrupt Buy Box management, a crucial feature that determines which seller's product is highlighted as the default option for purchase. When duplicates exist, multiple sellers offering the same product might unintentionally compete against themselves, splitting sales metrics and reducing their likelihood of winning the Buy Box. This can lead to unnecessary price wars and complicate marketplace analytics. Effective product deduplication ensures a clean catalog, allowing the Buy Box to function optimally by consolidating seller competition and attributing sales accurately. As a result, it enhances both operational efficiency and customer satisfaction. 

The challenge of product deduplication grows with the scale and diversity of product listings. In the Turkish e-commerce sector, this problem is particularly acute due to the rapid growth of the market and the high volume of unique sellers. Traditional keyword-based search methods fail to effectively identify duplicates because they rely on exact text matches rather than semantic similarities. For example, "wireless earbuds" and "Bluetooth earphones" might describe identical products but would be treated as different items by keyword-based systems. This limitation has led to the adoption of vector search-based solutions, which represent products as high-dimensional vectors for semantic similarity comparisons. While vector search methods excel at capturing nuanced relationships in product data, deploying such systems at scale in e-commerce introduces its own set of challenges, including inference latency, memory overhead, and maintaining efficiency in a growing catalog dataset with over 220 million products.

Existing solutions often leverage pretrained models due to their strong capabilities in generating meaningful representations. However, these models produce high-dimensional vectors. For example, our initial implementation using the Google Universal Sentence Encoder Multilingual \cite{cer2018universal} model generated 512-dimensional embeddings. While effective, it faced scalability issues when we hit indexing 8 million items, causing inference times to rise and operational costs to increase. Such limitations highlight the need for scalable, efficient models tailored to the specific requirements of e-commerce.

To address these bottlenecks in the vector search capabilities, we developed a domain-specific text model, fine-tuned for the linguistic and product-specific nuances of Turkish e-commerce,  which generates compact product representation. The model built on BERT \cite{kenton2019bert}, incorporating dimensionality reduction and enhanced feature extraction through intermediate transformer layers and convolutional layers. This approach enables us to compress representations to 128 dimensions without significant information loss. 

Additionally, we designed a decider model to classify product pairs as duplicates, leveraging text and image vectors to produce confidence scores. This classification-based architecture avoids the inefficiencies of pairwise comparisons inherent in siamese or contrastive learning, ensuring real-time inference and category-agnostic scalability. For the image representations that the decider model is used, we transitioned from traditional AutoEncoders to Masked AutoEncoders (MAE) with a structured patch selection logic, ensuring the preservation of critical visual details unique to e-commerce product images.

By integrating these advancements, our solution achieves improved scalability, efficiency, and accuracy in product deduplication. The proposed system addresses the unique challenges of the e-commerce domain, offering significant operational benefits and enhancing the customer shopping experience.

\section{Related Works}
Our work builds upon established research in feature extraction and multimodal learning, which we adapt for the specific operational constraints of a large-scale e-commerce environment.

\subsection{Text and Image Representation}
Transformer-based architectures, particularly BERT and its variants, have become the standard for semantic text representation by capturing deep bidirectional context \cite{kenton2019bert, sanh2019distilbert,jiang2020convbert}. We leverage BERTurk, a model pre-trained for the Turkish language, as our foundation \cite{stefan_schweter_2020_3770924}. Furthermore, the practice of adapting language models to specific domains is critical for capturing the nuances of e-commerce product titles, which often contain domain-specific jargon and structure.

For image understanding, the field has progressed from Convolutional Neural Networks (CNNs) to transformer-based approaches \cite{shaheen2016impact}. Unsupervised methods like Masked Autoencoders (MAEs) are particularly relevant, as they compel a model to learn contextual visual features rather than just surface-level patterns \cite{he2022masked}. This deeper understanding is crucial for accurately differentiating visually similar but distinct products.

\subsection{Foundational Multimodal Models and Our Approach}
The advent of large-scale Vision-Language Models (VLMs) like CLIP and BLIP-2 has enabled powerful joint representations of text and images. Trained on massive web-scale datasets, these models excel at general-purpose, zero-shot tasks by learning a broad, holistic understanding of visual and textual concepts \cite{radford2021learning,li2023blip}.

However, this generality comes at a significant operational cost. Their large parameter counts and high-dimensional vector outputs (e.g., 512 dimensions or higher) introduce prohibitive inference latency and memory requirements for real-time industrial applications like large-scale product deduplication. Our work diverges from this philosophy by prioritizing a domain-specific, efficiency-first approach. Instead of leveraging a single, large VLM, we design separate, lightweight models that generate compact, fast, and highly tailored embeddings to solve a targeted operational problem.

\subsection{Product Deduplication Approaches}
Matching similar products is an important part of a catalog management system in e-commerce. Duplicate products bloat the product catalog and may cause issues on both operational side and user experience. On the operations side, it increases costs due to increased or redundant manual labor for each product listing.  On the other hand, users face redundant search results and overall poor user experience. In order to alleviate these issues,various systems \cite{li2020deep,tracz-etal-2020-bert, peeters2022supervised, peeters2024entitymatchingusinglarge, herrero2024learning} are proposed.

\section{Theoretical Analysis}
In the e-commerce domain, product deduplication is a critical component of catalog management. Traditional keyword-based approaches often fall short due to inconsistencies in product titles and attribute usage. To address this, we implemented a vector search-based deduplication pipeline that generates robust representations of products using both textual and visual data.

The vector search retrieves the top-N candidate products using text-based vectors. However, identifying true duplicates from among these candidates proved non-trivial. Early attempts to rely on vector distances and rule-based filtering were insufficient due to the diverse ways sellers create product listings. This necessitated a dedicated classification model—referred to as the decider model—that could learn to identify duplicates using the semantic and visual context of the products.

While product images are essential for accurate comparison, generating visual vectors for all catalog items is computationally expensive. To optimize infrastructure use, image vectors were only employed in the decider model, whereas the initial vector search operated solely on text-based representations.

\section{Methodology}

Our multimodal product matching system was developed through several iterations, focusing on robust feature extraction, efficient dimensionality reduction, and an accurate decision model. This section details the individual steps that resulted in our final architecture.

\subsection{Feature Extraction and Dimensionality Reduction}

A central challenge in deploying deep models at scale is managing the high dimensionality of representation vectors. Pretrained models like BERTurk \cite{stefan_schweter_2020_3770924} and EfficientNetV2 \cite{DBLP:journals/corr/abs-2104-00298} yield 768- and 1792-dimensional vectors, respectively. These vectors are expensive to store, index, and search in large catalogs.

To determine the extent of feasible dimensionality reduction, we conducted a feasibility study. A binary decision model was trained using original and compressed versions of BERTurk and EfficientNetV2 representations. The original configuration achieved an F1 score of 0.86. Reducing both embeddings to 512 dimensions using PCA slightly changed the performance to 0.856. Further compression to 256 and 128 dimensions yielded F1 scores of 0.85 and 0.84, respectively. These results demonstrated that deduplication could be performed effectively using 128-dimensional vectors, which were adopted for all subsequent experiments to ensure scalability and efficiency.

Reducing dimensionality without compromising representation quality required more than simple projection. For textual data, we experimented with using token embeddings from intermediate BERT layers, which strike a balance between raw and highly contextualized representations. To further enhance efficiency, we added a CNN-based aggregation layer to automatically extract salient features from intermediate layers, improving performance while controlling vector size.

A similar approach was taken for images. Our initial solution involved using representations from an EfficientNet-based autoencoder. Encouraged by results from textual modeling, we later trained Masked AutoEncoders (MAEs), which preserve object-level contextual cues rather than superficial visual features. The initial approach of the model adapted only output dimensionality; the latest version employs guided patch selection to retain important visual details in the compressed representations.

\subsubsection{Text Feature Models}
We evaluated several open-source models to generate initial text representations. Among Turkish pretrained language models, as shown in Table \ref{tab:freq}, BERTurk achieved the best performance on our masked language modeling (MLM) task, outperforming ConvBERTurk and DistilBERTurk \cite{stefan_schweter_2020_3770924}.

\begin{table}[h!]
\caption{Text Model Variants MLM Validation Scores}
\label{tab:freq}
\centering
\begin{tabular}{|c|c|}
\hline
\textbf{Model Name}&\textbf{Validation Loss}\\
\hline
bertTurk & 1.51\\
\hline
convbertTurk & 1.89\\
\hline
distillbertTurk & 1.81\\
\hline
\end{tabular}
\end{table}

The initial text representation model employed BERT as a backbone, followed by a dimensionality reduction module (linear + regularization + activation) for compression and a classification module (linear + activation) for training. After training, the compressed “CLS” token served as the input sentence representation.

Building on this, a stricter feature extraction approach was adopted while maintaining the fixed 128-dimensional constraint. To achieve this, a compression algorithm was designed to refine the “CLS” token representation. Since using the “CLS” token in an MLM framework required a decoder, a simple restacking and shaping scheme was introduced.

\begin{figure}[!t]
\centering
\includegraphics[scale=0.25]{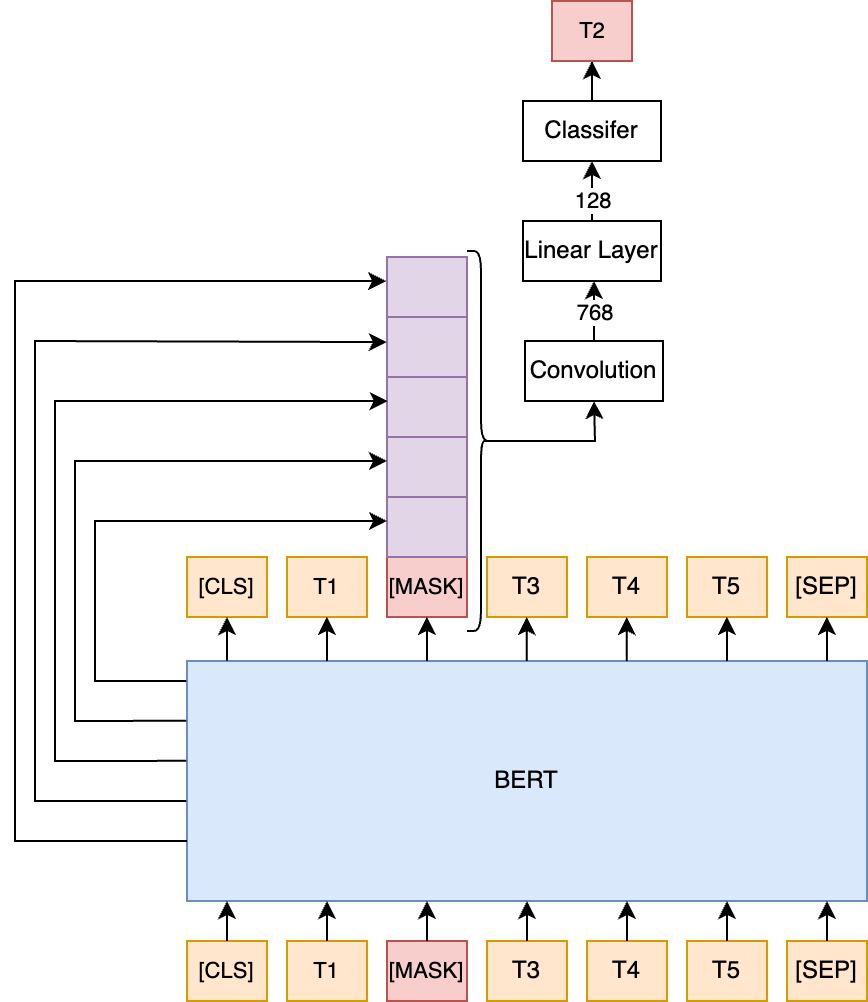}
\caption{Updated version of text model}
\label{fig:nlp_v2}
\end{figure}

To enhance information density and capture lower-level text features, intermediate BERT layer outputs are combined and passed to the compression step (see Fig. \ref{fig:nlp_v2}). Non-consecutive intermediate layers were selected, stacked, and processed through convolutional layers before being incorporated similarly to the original BERT output. This modification enabled the model to leverage early Transformer representations, improving its ability to extract richer e-commerce text features.

\subsubsection{Visual Feature Models}

For the image representations, DeiT \cite{DBLP:journals/corr/abs-2012-12877}, Swin Transformer \cite{DBLP:journals/corr/abs-2103-14030}, and LeViT \cite{DBLP:journals/corr/abs-2104-01136} were benchmarked on the masked autoencoding (MAE) task. DeiT achieved the highest Multi-scale Structural Similarity Index Measure (MS-SSIM) \cite{wang2003multiscale}, followed by Swin and LeViT. Results are shared in Table \ref{tab:image_backbone}.

\begin{figure*}[!t]
\centering
\includegraphics[scale=0.3]{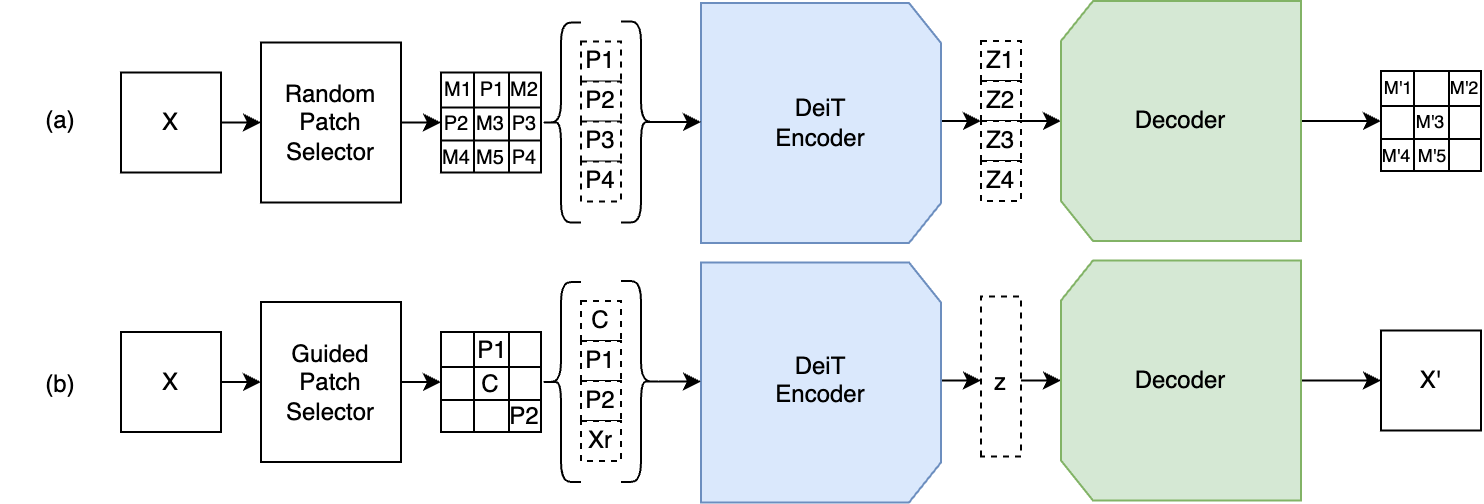}
\caption{(a) Second iteration of the image model \textit{P\#} denotes the selected patches, \textit{M\#} denotes the masked patches (b) Final iteration of the image model \textit{P\#} denotes the selected patches, \textit{C} is the center patch, \textit{Xr} is resized image}
\label{fig:image_v2v3}
\end{figure*}

\begin{table}[!h]
\caption{Image Model Variants MAE Validation Scores}
\label{tab:image_backbone}
\centering
\begin{tabular}{|c|c|}
\hline
\textbf{Model}& \textbf{MS-SSIM} \\
\hline
LeViT& 0.27 \\
\hline
Swin Transformer & 0.43 \\
\hline
DeiT& 0.45 \\
\hline
\end{tabular}
\end{table}

The initial approach for dense image representation extraction began with an AutoEncoder using an EfficientNetV2 encoder and a convolutional decoder. To maximize information density, no skip connections between encoder and decoder were made; the decoder relied solely on the compressed 128-length vector. However, analysis revealed a lack of contextual understanding: while similar objects in size and color had close representations, variants of the same object were mapped to distant vectors.

Exploring contextual image understanding, DeiT-based regular AutoEncoders were tested, but concerns over similar pitfalls led to adopting a Masked AutoEncoder (MAE) approach. Unlike traditional AutoEncoders, MAEs reconstruct masked image regions from unmasked patches, enforcing contextual learning (see Fig. \ref{fig:image_v2v3}-a).

In detail, standard MAEs divide images into “N” patches, mask a portion, and reconstruct the missing areas. Given our dimensional constraints, each unmasked patch was compressed to 32-length vectors, forming a total representation of 128. However, this traditional masking approach risked losing fine-grained details crucial in e-commerce images, such as product prints or brand markings. Additionally, as most image information is centrally located, randomly masking could result in samples missing key details.

To address this, we introduced a structured patch selection method (see Fig. \ref{fig:image_v2v3}-b):
\begin{itemize}
\item Divide the image into “k” equal patches.
\item Always select the center patch (containing the most information).
\item Select two random patches from the remaining eight.
\item Resize the full image to match a single patch size and use as the last patch.
\item Stack patches as standard MAE inputs.
\end{itemize}

This approach preserved some randomness while ensuring key details were captured. Unlike traditional MAEs predicting only masked regions, our model reconstructed the full image, improving its ability to differentiate subtle variations.

\subsubsection{Domain Adaptation}

Since pretrained models were trained on general-purpose datasets, domain adaptation was crucial. We investigated both supervised and unsupervised methods. Supervised training on leaf-category prediction was found to encourage excessive clustering within product categories, potentially masking distinctions between similar items (e.g., t-shirt and skirt in a single image).

Unsupervised approaches proved to be more effective. In both cases, the goal was to reconstruct the original inputs from compressed vectors, thereby forcing the models to preserve semantically important features. This reconstruction-based training helped improve the quality of the representations without relying on noisy or synthetic labels.

\subsection{Product Matching Pipeline}

Storing, continuously adding new, and updating product vectors necessitate the use of a dedicated vector database (VecDB) that ensures high availability. Conducting searches using a brute-force approach is resource-intensive; thus, most contemporary VecDB solutions incorporate faster approximate nearest neighbor (ANN) search capabilities through various indexing algorithms \cite{pan2024survey}.

For the VecDB setup, Milvus \cite{10.1145/3448016.3457550} was chosen for its active open-source community support, ease of setup, distributed capabilities, and its ability to execute various indexing and searching algorithms. The selection of indexing algorithms is crucial for optimizing the efficiency and speed of similarity searches. The two primary candidates considered were the clustering-based Inverted File FLAT (IVF\_FLAT) \cite{5432202} and the graph-based Hierarchical Navigable Small World (HNSW) \cite{8594636}. Quantized indexing algorithms were deliberately excluded to prevent unnecessary loss in recall rate. Initial tests with varying vector counts and dimensions across both algorithms indicated that, given our memory constraints, IVF\_FLAT emerged as the more suitable approach, provided that the appropriate cluster count and query units parameters are selected.

\begin{figure*}[!t]
\centering
\includegraphics[scale=0.33]{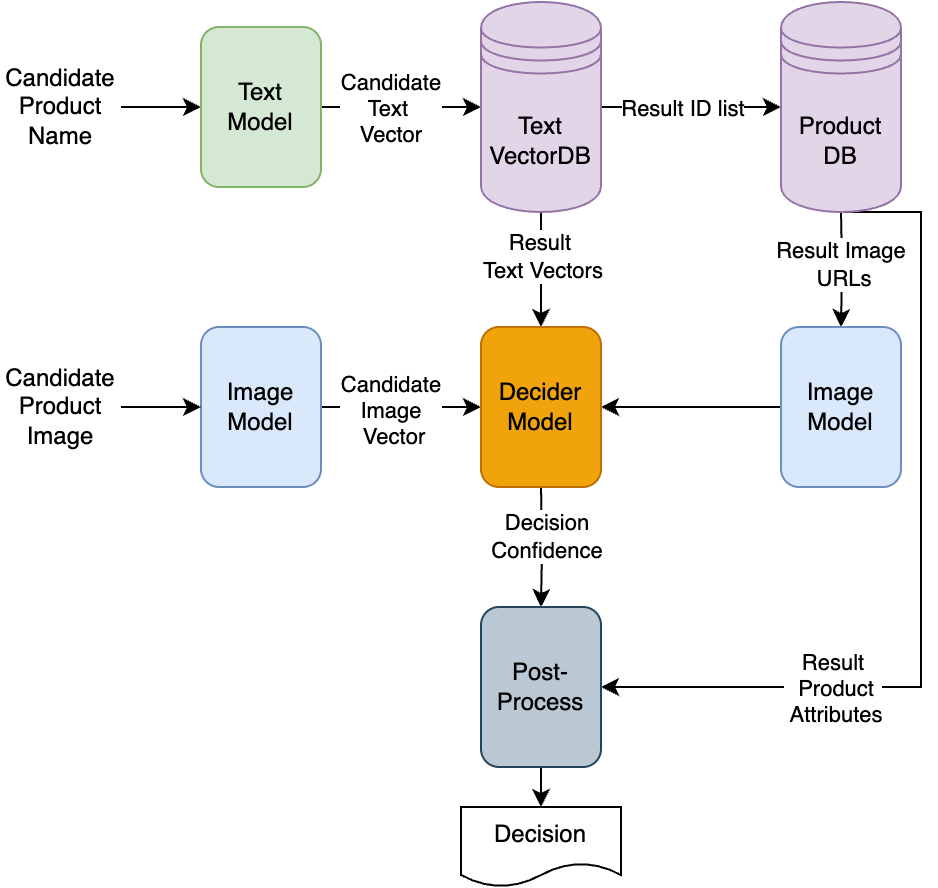}
\caption{Final system design for product deduplication.}
\label{fig:system1}
\end{figure*}

Development requirements for the decider model arose from the need to quickly determine if a pair of product listings belong to the same product. Initial trials using only vector distances showed promise; however, varying threshold values across different categories complicated management. To address this, a separate model was developed, leveraging dense vector representations to decide if given vectors correspond to the same product.

The decider model processes four vectors—two for image and two for text—associated with each product listing and generates a confidence score indicating the likelihood of them being the same product. A key architectural decision was to implement the model as a traditional classifier rather than using siamese or contrastive learning approaches. This choice was motivated by the need for efficient real-time inference, as standard classifiers can rapidly produce confidence scores from stored vectors. In contrast, siamese networks require computationally intensive pairwise comparisons, which are less scalable for large datasets.

Additionally, data availability and quality influenced this decision. Siamese and contrastive learning models typically rely on both paired and unpaired data to learn meaningful embeddings. However, only paired data was readily available, and generating synthetic unpaired data risked oversimplifying relationships, potentially hindering the model's ability to learn robust distinctions. Moreover, a classifier inherently handles multiple product categories within a unified framework, eliminating the need for category-specific thresholds and simplifying system scalability and management.

By choosing a conventional classification approach, the decider model achieves efficient inference, better aligns with available data, and reduces operational complexity. This ensures high performance and robustness across diverse product categories, making the model well-suited for real-world applications in product listing management. The final system design can be seen in Fig. \ref{fig:system1}.

\section{Experiments \& Results}

To validate the effectiveness and efficiency of our proposed multimodal deduplication system, we conducted a series of experiments on a large-scale, real-world e-commerce dataset. We evaluated our model's performance against a strong commercial baseline and analyzed its operational efficiency in terms of inference speed and memory consumption. Furthermore, we present ablation studies to justify our key architectural decisions, including our compact embedding strategy and text feature aggregation method. This section details the experimental setup, presents the main performance results, and concludes with an error analysis to identify areas for future improvement.

\subsection{Experimental Setup}

This section details the experimental setup used to evaluate our system, including the dataset composition, the baseline model for comparison, and the metrics for performance assessment.

\subsubsection{Dataset} All models were trained and evaluated on a curated internal dataset from the Hepsiburada e-commerce catalog, consisting of 13 million products across 4,500 different categories. The feature extraction models were pre-trained on this large corpus of product titles and their corresponding primary images. The decider model was trained on a set of 350,000 manually-labeled product pairs and tested on a hold-out set of 8,000 pairs, which were labeled by internal catalog experts. While this internal dataset ensures high domain relevance, we acknowledge that evaluation on public e-commerce datasets would further strengthen the generalizability of our findings.

\subsubsection{Baseline} We compare our system against a black-box, commercially available API-based deduplication service, referred to as "3rd Party." This represents a strong, production-level alternative.

\subsubsection{Metrics} We report Precision, Recall, and F1-score for both "Match" and "Not Match" classes. The primary evaluation metric is the macro-average F1-score, which provides a balanced measure of performance across both classes.

\subsubsection{Training Procedure} For both text and image models, we utilized the AdamW optimizer with a starting learning rate of 0.0001, complemented by a cosine scheduler that scales the rate down toward zero. This setup proved most efficient, resulting in the lowest loss values during training until convergence.

\subsection{Main Results}
The primary evaluation of our system's performance is presented in Table \ref{tab:test_results}. Our model achieves a macro-average F1 score of 0.90, demonstrating a significant and robust improvement over the third-party solution's score of 0.83. A deeper analysis of the per-class metrics reveals the practical advantages of our approach for the e-commerce domain.

\begin{table}[!h]
\centering
\caption{Performance Comparison Results}
\label{tab:test_results}
\begin{tabular}{|c|c|c|c|c|}
\hline
\textbf{Model} & \textbf{Class} & \textbf{Precision} & \textbf{Recall} & \textbf{F1}  \\
\hline
Internal       & Match          & 0.91               & 0.85            & 0.88         \\
\hline
3rd Party      & Match          & 0.90               & 0.73            & 0.81         \\
\hline
Internal       & Not Match      & 0.88               & 0.96            & 0.92         \\
\hline
3rd Party      & Not Match      & 0.76               & 0.94            & 0.84         \\
\hline
\end{tabular}
\end{table}

The most critical improvement is in the "Match" class, where our system achieves a Recall of 0.85, compared to 0.73 for the baseline. This 12-point increase is highly significant, as it indicates that our system is substantially more effective at its core task: identifying true duplicate products. This directly translates to a cleaner product catalog, reduced customer confusion, and more effective Buy Box management. Although our precision is marginally higher (0.91 vs. 0.90), the substantial gain in recall underscores our model's superior ability to find duplicates that the other system misses.
For the "Not Match" class, our system maintains excellent performance with a Recall of 0.96 and Precision of 0.88. This demonstrates that our model is highly reliable in distinguishing between different products, minimizing false positives that would otherwise require costly manual review and erode trust in the automated system.

It is important to note the methodology used to generate the test data. The 8,000 labeled pairs were curated from candidates flagged by our system and the third-party solution. Although this introduces a known bias—focusing on ambiguous cases where both systems are active—it serves as a necessary and pragmatic starting point for a direct, head-to-head comparison. It effectively evaluates how each system performs on a challenging set of real-world examples, allowing us to identify clear areas for improvement and demonstrate our system's superior performance on this competitive ground.

\subsection{System Efficiency and Architectural Analysis}
To validate the system's practicality for industrial applications and to justify our key design choices, we analyzed its efficiency and the architectural decisions that led to the final model.

\subsubsection{System Efficiency and Scalability}
A core design principle of our system was operational efficiency, ensuring that it could scale cost-effectively to hundreds of millions of products. Our experimental results validate this approach. As detailed in Table \ref{tab:milvus_results}, the memory footprint for indexing 10 million of our 128-dimensional vectors is a manageable 5.5 GB. This represents a four-fold reduction compared to the 20+ GB that would be needed for more standard 512-dimensional embeddings, a critical advantage for large-scale deployment. 

\begin{table}[!h]
\centering
\caption{RAM Usage of 10 Million Vectors on Milvus}
\label{tab:milvus_results}
\begin{tabular}{|c|c|}
\hline
\textbf{Vector Size} & \textbf{RAM Usage (GB)}  \\
\hline
128                  & 5.5                         \\
\hline
256                  & 10.2                        \\
\hline
512                  & 20                          \\
\hline
1024                 & 39.7                        \\
\hline
\end{tabular}
\end{table}

Beyond memory, the system is optimized for low-latency inference. The results in Table \ref{tab:inference_results} show that our model architecture is more thant 6 times faster than CLIP and 30 times faster than BLIP when processing 1,000 product pairs. Together, these quantitative gains in memory and speed confirm that our system is not only accurate but also robust and efficient enough for a high-throughput production environment.

\begin{table}[!h]
\centering
\caption{Inference Times per 1000 Product Pairs}
\label{tab:inference_results}
\begin{tabular}{|c|c|}
\hline
\textbf{Model} & \textbf{Generation Time (sec)} \\
\hline
Ours  & 10                    \\
\hline
CLIP  & 62                    \\
\hline
BLIP  & 302                   \\
\hline
\end{tabular}
\end{table}

\subsubsection{Architectural Analysis of the Text Model} 
Our design process for the text model was data-driven. A preliminary analysis of 13 million product titles revealed that the most critical information is typically front-loaded, with the first 32 tokens capturing the essential details for over 95\% of listings. This made a fixed 32-token input a feasible approach.

Our initial text model used a standard BERTurk backbone, where the final [CLS] token was compressed to 128 dimensions. This base model, trained on a Masked Language Modeling (MLM) task, achieved a validation loss of 1.51. However, we hypothesized that relying solely on the final, highly contextualized layer might discard valuable lower-level information from earlier layers.
To capture a richer representation, we enhanced the architecture. We extracted the outputs from intermediate transformer layers—specifically layers 1, 3, 5, 7, 9, and 11—and stacked them with the output from the final layer. This multi-layer representation was then passed through a 1D convolutional chain followed by normalization and activation. This allowed the model to selectively learn the most salient features across different levels of contextualization. This enhanced model achieved a significantly lower MLM validation loss of 1.43, confirming its superior ability to capture relevant token-level information. The final [CLS] token from this improved architecture was then used as the 128-dimensional output for all downstream tasks.

\subsubsection{Architectural Analysis of the Image Model}
The development of the image model began with a feasibility study to determine the viability of compact 128-dimensional embeddings. We trained a binary decision model using original high-dimensional representations from BERTurk and EfficientNetV2, which achieved an F1-score of 0.86. We then compressed these embeddings using PCA to various dimensions. The performance drop was minimal: 512 dimensions yielded an F1 of 0.856, 256 dimensions an F1 of 0.85, and 128 dimensions an F1 of 0.84. This minor 2\% drop in F1-score for a 128-dimensional vector was a clear and acceptable trade-off for the immense gains in system efficiency.

Our first implementation was a simple AutoEncoder with an EfficientNetV2 encoder and a convolutional decoder, using Multi-scale Structural Similarity (MS-SSIM) as the loss function to avoid blurriness issues common with MSE. However, this model lacked contextual understanding; while images with similar colors and shapes had close vector representations, different variants of the same object were often mapped to distant vectors.

This led us to adopt a Masked Autoencoder (MAE) approach with a DeiT backbone to force contextual learning. To adapt MAE for e-commerce, where central details like brand logos are critical, we replaced random masking with a structured patch selection logic. For each image, we divide it into 9 patches and create a 4-patch input sequence consisting of: (1) the center patch, to preserve key details, (2) two random patches, to retain contextual understanding, and (3) a resized version of the entire image, to capture the overall object shape. The model is trained to reconstruct the full image from the resulting compressed 128-dimensional vector produced by the encoder.

\subsubsection{Robustness to Image Scaling} 

\begin{figure}[h]
  \centering
  \includegraphics[width=\linewidth]{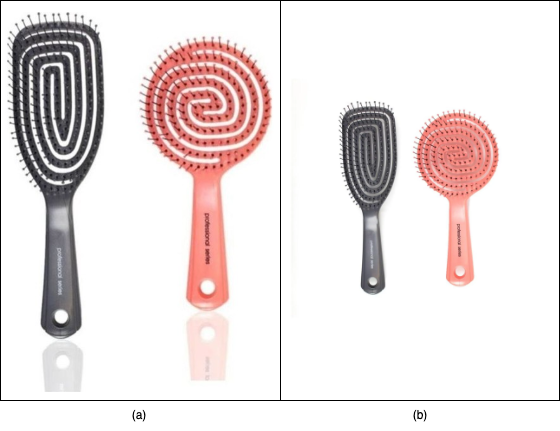}
  \caption{Scale differences of a product. (a) image does not have a white border, (b) image has a white border}
  \label{fig:im_scale_ex} 
\end{figure}

During early matching trials, we encountered a persistent issue where variations in object scale and image padding (as shown in Fig.  \ref{fig:im_scale_ex}) caused the vectors for the same product to be dissimilar. An initial plan to use a separate preprocessing module to remove these "boring pixels" proved too computationally expensive for our production environment. Instead, we integrated this logic as a data augmentation step directly into the dataloader. A random number determines if the augmentation is applied. If so, the algorithm converts the image to black and white, finds the coordinates of the first non-zero pixels to form a bounding box around the object, and returns this cropped image as the model input. This process forced the model to learn scale-invariant representations, significantly improving its robustness to the diverse image styles submitted by different sellers.

\subsection{Error Analysis}
A qualitative analysis of the model's failures revealed a primary weakness: differentiating between products with fine-grained visual distinctions. Notably, the model showed a propensity to classify visually similar but distinct product variants as matches. For instance, products differing only in subtle packaging variations or minor design elements, while fundamentally representing different SKUs, were sometimes grouped together. This suggests that while our multimodal embeddings excel at capturing broad semantic and visual similarities, they occasionally struggle with fine-grained distinctions crucial for differentiating product variants that are visually very close but operationally separate. Addressing these nuances will be a key focus for future refinements.

\section{Conclusion}
By strategically combining unsupervised learning with domain-specific adaptations, we developed an efficient and scalable product deduplication method. Compressing vector dimensions while retaining critical semantic information enabled accurate deduplication with reduced storage and computational costs. Our results demonstrate that well-designed compact vectors can effectively capture product listing nuances, facilitating reliable large-scale deduplication.

Integrating multimodal vector representations, combining text and image embeddings, further improved duplicate detection accuracy. Our decider model, leveraging these embeddings, achieved a macro-average F1 score of 0.90, highlighting its robustness. Additionally, the use of Milvus ensures efficient similarity searches in expanding datasets, enhancing operational efficiency and customer experience. Deployed architecture is serving our internal users with 116M\footnote{as of August 2025} product vectors daily.

This study underscores the potential of domain-specific machine learning adaptations in e-commerce. Future work will focus on developing a unified multimodal framework that integrates text and image features to enhance matching accuracy. Efforts will also be directed toward creating lighter and faster models to improve scalability in real-world applications. Furthermore, to address issues arising from product variants and the insufficiency of manual attribute controls, we plan to integrate product attribute information into the decider model. Additionally, future work will involve extending the application to other languages. These advancements aim to create a more accurate, efficient, and adaptable product matching system capable of addressing the dynamic challenges of large-scale e-commerce.

\bibliographystyle{IEEEtran}
\bibliography{IEEEexample.bib}

\end{document}